\documentclass[aps,prl,twocolumn,nofootinbib]{revtex4-1}

\usepackage{url,comment}
\usepackage{times}
\usepackage{latexsym}
\usepackage{graphicx, graphics, hyperref, amsmath, amssymb, slashed, xcolor, bbm,amsthm, array}
 \usepackage{subfigure}

\usepackage{rotating}
\usepackage{afterpage}

\newcommand{\nc}{\newcommand}
\nc{\beq}{\begin{equation}}
\nc{\eeq}{\end{equation}}
\nc{\barray}{\begin{eqnarray}}
\nc{\earray}{\end{eqnarray}}
\nc{\barrayn}{\begin{eqnarray*}}
\nc{\earrayn}{\end{eqnarray*}}
\nc{\bcenter}{\begin{center}}
\nc{\ecenter}{\end{center}}
\nc{\mc}{\mathcal}
\nc{\er}[1]{(\ref{eq:#1})}
\nc{\onehalf}{\frac{1}{2}} 
\nc{\partialbar}{\bar{\partial}}
\nc{\psit}{\widetilde{\psi}}
\nc{\Tr}{\mbox{Tr}}
\nc{\hc}{\mbox{H.c.}}
\nc{\ev}{\;\mathrm{eV}}
\nc{\mev}{\;\mathrm{MeV}}
\nc{\gev}{\;\mathrm{GeV}}
\nc{\tev}{\;\mathrm{TeV}}

\def\chii0{\chi_i^0}
\def\chij0{\chi_j^0}

\newcommand{\gsim}{\lower.7ex\hbox{$\;\stackrel{\textstyle>}{\sim}\;$}}
\newcommand{\lsim}{\lower.7ex\hbox{$\;\stackrel{\textstyle<}{\sim}\;$}}
\nc{\ttbar}{t\bar t}
\def\ifb{{\ \rm fb}^{-1}}

\newcommand{\fref}[1]{Fig.~\ref{f.#1}}
\newcommand{\eref}[1]{Eq.~(\ref{e.#1})}

\newcommand{\cref}[1]{Chapter~\ref{c.#1}}

\graphicspath{{plots/}}

\begin{document}

\title{
A Quirky Probe of Neutral Naturalness
}

\author{Zackaria Chacko}
\email{zchacko@umd.edu}
\author{David Curtin}
\email{dcurtin1@umd.edu}
\author{Christopher B. Verhaaren}
\email{cver@umd.edu}

\affiliation{Maryland Center for Fundamental Physics, Department of Physics,\\ University of Maryland, College Park, MD 20742-4111 USA}

\date{\today}
\begin{abstract}
We consider the signals arising from top partner pair production at the LHC as a probe of theories of Neutral Naturalness. 
We focus on scenarios in which  top partners carry electroweak charges, such as Folded SUSY or the Quirky Little Higgs. 
In this class of theories the top partners are pair produced as quirky bound states, since they are charged under a mirror color group whose lightest states are hidden glueballs. The quirks promptly de-excite and annihilate into glueballs, which decay back to SM fermions via Higgs mixing. This can give rise to spectacular signatures at the LHC, such displaced decays, or high-multiplicity prompt production of many hard $\bar b b$ or $\tau^+ \tau^-$ pairs. 
We show that signals arising from top partner pair production constitute the primary discovery channel for this class of theories in most regions of parameter space, and might provide the only experimental probe of scenarios with sub-cm glueball decay lengths. The measurement of top partner masses and couplings, which could be used to test the neutral naturalness mechanism directly, is also a tantalizing possibility.
\end{abstract}

%


%

\pacs{}%

\keywords{}

\maketitle

The Standard Model (SM) 
is a theoretical triumph whose final component, 
 a Higgs boson with approximately the expected couplings, was discovered 
  at the Large Hadron Collider (LHC) in 2012 \cite{Aad:2012tfa,Chatrchyan:2012ufa}. Despite its experimental successes, it suffers from a 
  \emph{hierarchy problem} \cite{Weisskopf:1939zz}: 
 quadratically divergent quantum corrections to the Higgs mass parameter 
 must 
  cancel against the bare mass term to obtain the measured 125 GeV mass. 
  From a Wilsonian Effective Field Theory viewpoint, 
  we 
   expect new degrees of freedom to couple to the Higgs and cancel the SM loops. 
   The largest 
   divergence, 
   from the top quark, 
  implies new physics 
  at a scale below $\sim \tev$. Otherwise, the theory is \emph{tuned} or \emph{unnatural}.

Theories like 
 supersymmetry \cite{Martin:1997ns} or the Little Higgs \cite{ArkaniHamed:2001nc, ArkaniHamed:2002qx, ArkaniHamed:2002qy, Schmaltz:2004de} cancel the top loop 
  by \emph{top partners} that are related to the top by a symmetry transformation. The symmetry relates the Higgs couplings of the top and top partner, 
  enforcing the cancelation. These top partners carry SM color, 
   leading to copious production at the LHC for 
 masses 
  below the TeV scale. While the absence of such a discovery at the first run of the LHC can be explained by kinematic blind-spots or non-minimal scenarios \cite{Martin:2007gf, Martin:2008sv,LeCompte:2011fh,Belanger:2012mk,Rolbiecki:2012gn,Curtin:2014zua,Kim:2014eva,Czakon:2014fka,CMS:2014exa, Rolbiecki:2015lsa, An:2015uwa}, these null results lead to some tension with naturalness.

In theories of \emph{Neutral Naturalness} (NN) \cite{Chacko:2005pe, Burdman:2006tz, Cai:2008au} the top loop is canceled by 
top partners without 
 SM color charge. This can occur when the symmetry that protects the Higgs mass does not commute with SM color. Such theories are clearly 
  consistent with LHC limits on colored particles. They also offer a more general framework for considering 
the 
experimental consequences of naturalness. The phenomenology of these models can be very rich, and, 
in general, radically different from 
 colored top partner scenarios.

Usually, NN 
top partners 
are charged under a mirror copy of QCD. They may 
carry SM electroweak (EW) charge, as in the case of Folded Supersymmetry (FSUSY) \cite{Burdman:2006tz} and the Quirky Little Higgs (QLH) \cite{Cai:2008au}, or remain SM singlets, as in the Twin Higgs (TH) \cite{Chacko:2005pe, Barbieri:2005ri, Chacko:2005vw} family of theories.  These models have rich implications for cosmology \cite{
Garcia:2015loa, 
Craig:2015xla, 
Garcia:2015toa, 
Farina:2015uea, 
Schwaller:2015tja, 
Poland:2008ev, 
Batell:2015aha}, and possibly 
 flavor \cite{Csaki:2015gfd}. UV completions \cite{Craig:2014aea, Craig:2014roa, Batra:2008jy,Barbieri:2015lqa, Low:2015nqa, Geller:2014kta,
Craig:2013fga, Craig:2014fka, Chang:2006ra} are required at scales of order $5 - 10 \tev$ to protect against higher loop effects. At these energies the full protection mechanism of the theory is expected to become apparent. This strongly motivates the construction of future lepton and 100 TeV colliders \cite{Curtin:2015bka, Cheng:2015buv}.

At the LHC, 
the most promising signals of NN are \emph{displaced signatures} that arise when these theories realize a specific Hidden 
Valley \cite{Strassler:2006im, Strassler:2006ri, Strassler:2006qa, Han:2007ae} scenario. This was first explicitly pointed out in 
the context of the Fraternal Twin Higgs model~\cite{Craig:2015pha}. Without 
light matter charged under 
mirror color, 
 the lightest hidden hadrons 
are glueballs~\cite{Morningstar:1999rf}. Mirror gluons couple to the Higgs via a dimension-6 operator generated by the top partner loop 
\cite{Juknevich:2009ji}, similar 
to SM tops and gluons. This operator generates mixing between the $0^{++}$ glueball and the Higgs, allowing 
these states to decay to SM particles, primarily $\bar b b$ and $\tau^+ \tau^-$. 
 These decays are slow on collider timescales, with 
characteristic decay lengths of $\mu$m - km, which are reconstructable 
 in 
  LHC detectors. 
 
Glueball signals are particularly motivated for EW-charged top partners since 
 LEP constraints~\cite{pdg} forbid light mirror matter. 
Naturalness motivates top partner masses below a few TeV.
Renormalization group arguments then motivate $0^{++}$ masses between 
 10 and 60 GeV, 
  allowing for exotic Higgs decays. Displaced searches at the LHC for mirror glueballs arising from Higgs decays are 
 projected to be sensitive to 600 - 800 GeV top partners at the end of run 2, and TeV-scale top partners by the end of the HL-LHC \cite{Curtin:2015fna},  see 
  \fref{qlh}. Even the first $20 \ifb$ of 13 TeV data offer a reach of a few hundred GeV \cite{Csaki:2015fba}. By comparison, precision measurements of $h\to \gamma \gamma$ will only probe 
  top partner masses of a few 100 GeV \cite{Burdman:2014zta}.  This 
  illustrate the exquisite sensitivity of exotic Higgs decays to new physics 
\cite{Curtin:2013fra}, but large uncertainties remain. Most significantly, it is currently unknown how well hadronic sub-cm macroscopic decay lengths can be reconstructed and distinguished from background at the LHC. In \fref{qlh}, the orange regions that are not covered by the blue regions have relatively short-lived glueball decays, and it is not clear if displaced searches can be conducted with little background. Alternative probes of this \emph{sub-cm glueball decay} regime are highly motivated.

This letter 
 investigates another promising avenue for probing 
 NN: glueball signatures 
 from \emph{direct top partner production}. In theories such as FSUSY and QLH, top partners can be pair 
produced 
with sizable rate at the LHC. These 
pairs form a 
\emph{quirky bound state} \cite{Okun:1980kw,Okun:1980mu,Kang:2008ea}, 
since the mirror gluon string connecting them cannot snap by exciting 
light quark pairs out of the vacuum. These quirks can annihilate to mirror gluon jets.
 The glueballs resulting from mirror hadronization can then give rise to 
events with multiple displaced vertices, or multiple $\bar b b$ and $\tau^+ \tau^-$ pairs if the glueball decay is relatively short-lived. 
Quirky 
pair production also offers the 
possibility of measuring 
top partner masses and couplings directly, which could confirm
 the NN solution to 
  the Little Hierarchy Problem.

Glueballs produced from top partner annihilation 
 generally have higher multiplicity and momentum than those from 
 exotic Higgs decays. The overall production cross section can also be higher. For glueballs with long lifetimes, this means that direct top partner production could be discovered before exotic Higgs decays. On the other hand, depending on reconstruction efficiencies and backgrounds for displaced decays, top partner pair production may provide the only experimental probe of the sub-cm glueball regime, since the additional boost increases 
  decay length, and even ``prompt'' glueballs decaying to $\bar b b$ or $\tau^+ \tau^-$ will be discovered if their multiplicity and momentum 
   are high. By contrast, exotic Higgs decays to $4b$ are very difficult to discover without additional handles like displaced decays \cite{Curtin:2013fra}.

Quirky signals of FSUSY were 
considered in \cite{Burdman:2008ek}. However, they 
 focused on pair production of first and second generation partners and annihilation into $W\gamma$. The masses of those states are not as closely connected to naturalness as 
 the top partners, and this final state has much more SM background than displaced decays or high-multiplicity glueball final states. 

We show that pair production of top partners which annihilate 
into mirror glueballs, is 
 the discovery signature of NN at the LHC in large regions of parameter space, and provides an alternative 
  probe of the sub-cm glueball regime. A key challenge in making this prediction is the quantitative treatment of mirror hadronization, which is 
  not well understood in 
  pure $SU(3)$ gauge theory. Even so, we demonstrate how to consistently parameterize 
  ignorance of 
  the non-perturbative physics in 
   the hidden sector, and 
    systematically study 
     the signatures. We identify 
      regions of parameter space in which direct production is definitely superior to exotic Higgs decays as a probe of top partner mass, even with 
      pessimistic assumptions about the hadronization of the mirror gluon jets. 
A full exploration of the signature space, which can include final states with many $b \bar b$ pairs, displaced vertices, and missing energy, and which might allow for the measurement of top partner masses and couplings, will be explored in a detailed follow-up publication \cite{futurequirk}.

\begin{figure}
\begin{center}
\includegraphics[width=0.43\textwidth]{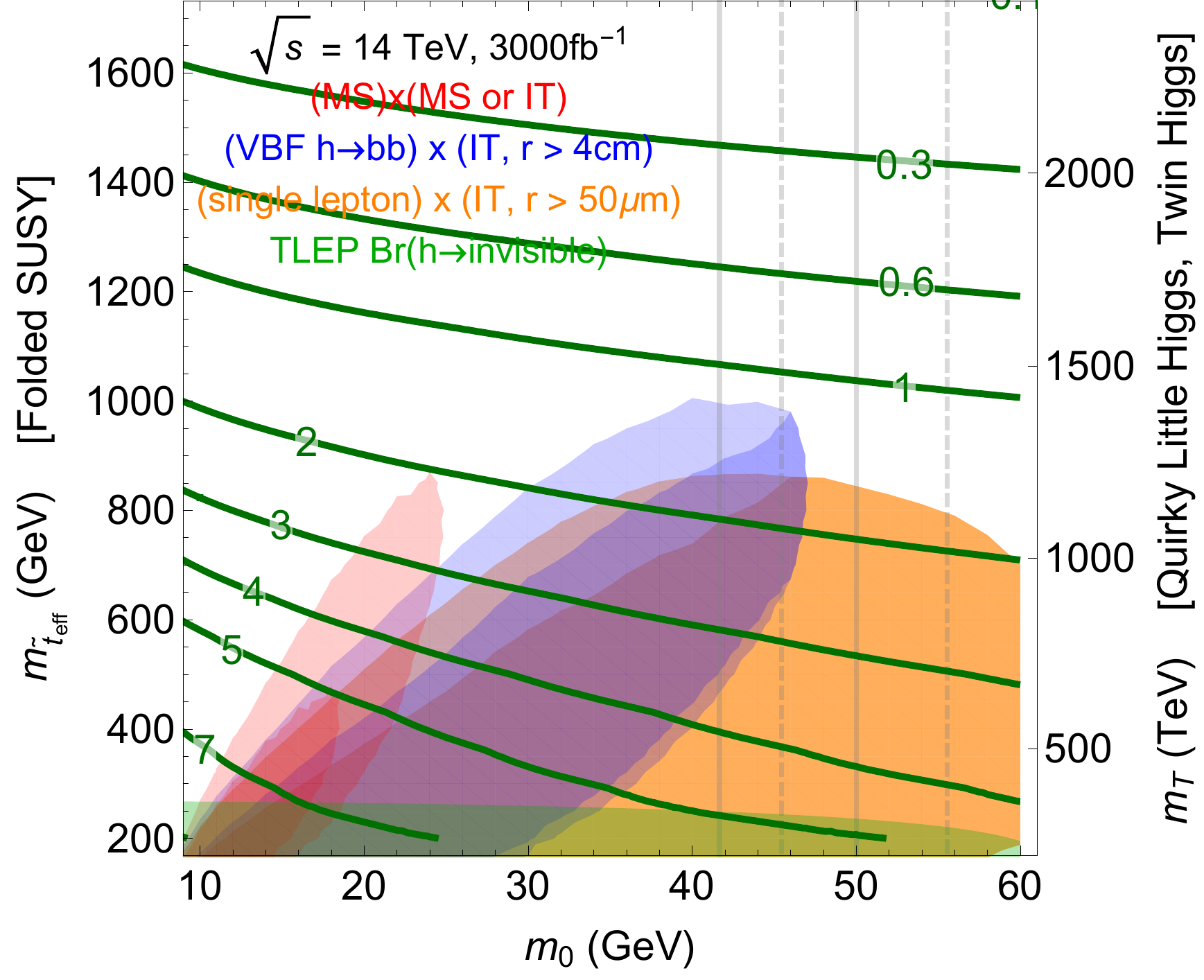}
\end{center}
\caption{\label{f.qlh}
\emph{Shaded regions: }
Projected sensitivity of 
 displaced vertex searches, at the 14 TeV LHC with $3000 \ifb$, to 
 mirror glueballs from exotic Higgs decays in theories of Neutral Naturalness \cite{Curtin:2015fna}. 
 Bounds are expressed as a function of lightest glueball mass $m_0$ 
  and top partner mass, $m_{\tilde t_\mathrm{eff}}$ in FSUSY for degenerate unmixed stops (left axis) and $m_T$ in  TH/QLH (right axis). Light shading represents the factor of $\sim 10$ uncertainty in the number of 
  $0^{++}$ glueballs produced during mirror hadronization. 
    \emph{Green contours}: 
   Conservative estimate of the number of glueballs produced from top partner pair production and annihilation in the QLH model, normalized to the rate from exotic Higgs decays, see \eref{Rqlh}.
}
\end{figure}


We now analyze top partner pair production in FSUSY and the QLH. These models serve as useful 
benchmarks, but our conclusions are general and should apply to all 
 EW-charged top partners charged under a mirror QCD force.


\textbf{FOLDED SUPERSYMMETRY --- }
In the 5D FSUSY theory \cite{Burdman:2006tz}, all QCD-charged fields of the MSSM, and the $SU(3)_c$ gauge sector itself, are duplicated into two sectors $A$ (SM) and $B$ (mirror) at some multi-TeV scale, with couplings related by a discrete $\mathbb{Z}_2$ symmetry. At energies $\lesssim \tev$, the electroweak and Higgs sectors are similar to the 4D MSSM with decoupled gauginos. However, only the 
$A$-sector quarks and $B$-sector 
 squarks have 
  zero modes. This realizes an accidental low-energy SUSY limit, with 
  quadratically divergent top contributions to the Higgs mass 
  cancelled by 
  mirror-sector stops, which are identical to conventional stops, except they 
   are charged under 
   mirror QCD.

For our purposes, 
the expressions for the lightest squark masses in FSUSY 
can 
be taken to be those of 
the MSSM \cite{Martin:1997ns}.
The light mirror hadrons are glueballs, as described above. Following the methodology of \cite{Curtin:2015fna}, we concentrate on the signatures of the 
$0^{++}$ glueball.
The stop masses and mixing angle $\theta_t$ are 
free parameters.

%
%
%

 The stops are produced electroweakly, with a cross section that is 
 readily computed 
 in MadGraph \cite{Alwall:2014hca}. 
They then 
 form a quirky bound state, connected by a 
 flux tube that is unbreakable in 
 the absence of light mirror QCD-charged matter. The bound state sheds energy by emitting 
 soft glueballs and photons, 
with the non-relativistic stops forming $s$-wave stoponium  $\eta_{\tilde{t}}$ before annihilating~\cite{Kang:2008ea}. The annihilation branching fractions are 
adapted from 
\cite{Martin:2008sv}.
Because of the large hidden sector QCD coupling and gluon multiplicity, the mirror di-gluon final state usually dominates, with a branching ratio of $\sim 50 - 80\%$ in most of our parameter space of interest. 
For large stop mass splittings and mixings, however, annihilation to two $125 \gev$ Higgs bosons can dominate (see also \cite{Batell:2015zla}), while $WW$, $ZZ$ are produced $\sim 10\%$ of the time, and $\gamma \gamma$ has $\mathcal{O}(10^{-3})$ branching fraction.  These SM final states may be particularly useful for precise mass measurements. 
Here we focus on the mirror gluon final state due to the low background of displaced searches.

If lighter states (like the sbottom) are available, one or both of the stops may $\beta$-decay, adding 
leptons to the mirror gluon jet signature. Whether $\beta$ decay occurs before annihilation depends on the mass-splitting \cite{Burdman:2008ek,Harnik:2008ax}.  We concentrate on the case where the lightest stop is pair produced and cannot $\beta$ decay, and indicate 
where this may not hold.

Our conservative estimate 
ignores the soft emission of photons and glueballs during de-excitation, 
concentrating 
on the mirror gluon jets 
created when the quirk state annihilates.

\textbf{\emph{Mirror Gluon Jets ---}}
The perturbative showering 
of the mirror gluons proceeds very similarly to the SM, except without quarks and with a coupling $\alpha_s^B$ that is a modest $\mathcal{O}(1)$ factor higher than the SM $\alpha_s^A$ due to differences in RG evolution \cite{Curtin:2015fna}. This makes the mirror jets pencil-like, with similar or slightly larger width than in the SM.

Next, we need to know how many 
 glueballs are produced in each jet (which determines glueball momentum), and what fraction are the 
  $0^{++}$ that give rise to displaced vertices. Unfortunately, the details of pure gauge hadronization, and how to reliably calculate them, are completely unknown. 
  Therefore, we parameterize 
  our ignorance such that we can 
  systematically consider the range of hadronization possibilities. 
  Our aim is parametric transparency and accuracy  
   with $\mathcal{O}(1)$ precision for the overall signal estimate, while factorizing from the ``hard'' theory parameters like top partner and glueball masses. 

Glueball multiplicities are encoded in the nonperturbative fragmentation function of the mirror gluon. While its magnitude is unknown, 
the DGLAP 
equation~\cite{Ellis:1991qj} determines 
how it changes with scale. 
In the massless limit, hadron multiplicities scale as
\begin{equation}
\langle n(E_{\text{CM}}^2)\rangle \propto \exp\left(\frac{12\pi}{33}\sqrt{\frac{6}{\pi\alpha_s^B(E_\text{CM}^2)}}+\frac14 \ln\alpha_s^B(E_\text{CM}^2) \right),\label{e.HadMult}
\end{equation}
where $\alpha_s^B(E_\mathrm{CM})$ is determined by the glueball mass (which fixes $\Lambda_\mathrm{QCD}^B$) and the assumption that the stop is the lightest mirror-QCD charged particle. 

Therefore, we define $N_\mathrm{G}(E_\mathrm{CM})$ as the \emph{total} number of glueballs produced, on average, by mirror gluon 
hadronization. 
Its dependence on the  center-of-mass energy $E_\mathrm{CM} \gtrsim 2 m_{\tilde t}$ is given by \eref{HadMult}, and fixed for all events and 
 stop masses once $N_G$ is specified at a given $E_\mathrm{CM}$. We also define $r_{G_0}$ as the fraction of those glueballs that are the lightest $G_0 = 0^{++}$ state. 

Thus, we encapsulate our ignorance of mirror hadronization by considering the 
parameter space of possible values 
$(N_G^0, r_{G_0})$, where $N_G^0 = N_G(E_\mathrm{CM})$ for some fixed $E_\mathrm{CM}$. This space is bounded: $N_G^0 \geq 1$ but smaller (per degree of freedom) than 
charged hadron production in the SM, since glueballs are heavier and more expensive to produce. (There is also an upper bound for light stops due to the non-negligible mass of mirror glueballs.)  Similarly, $ r_{G_0} \leq 1$ and likely larger than 
0.1, and has been estimated 
to be $\sim 0.5$ \cite{JuknevichPhD}.

\begin{figure}
\begin{center}
\begin{tabular}{cc}
\includegraphics[width=0.43\textwidth]{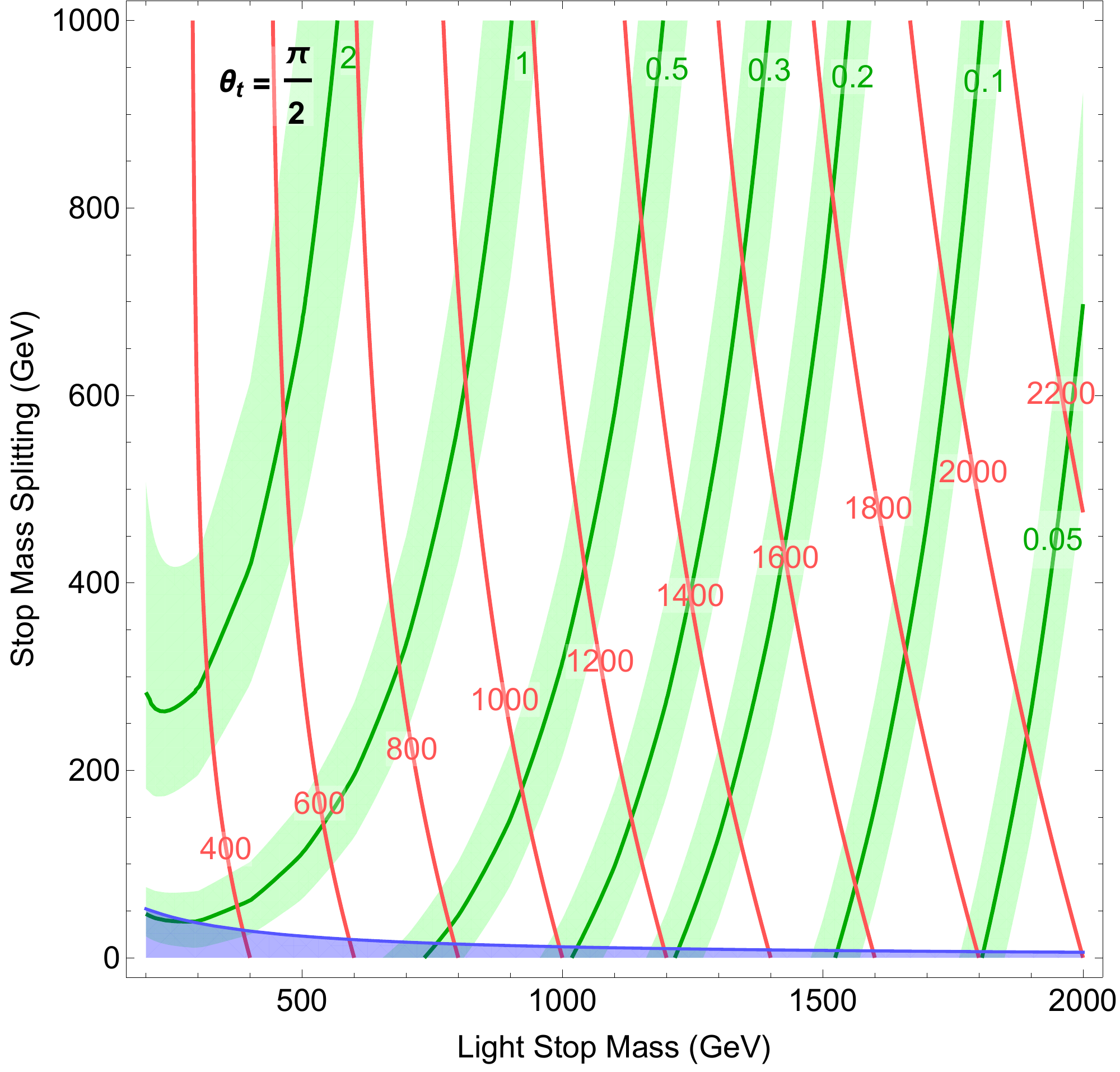}
\\
\includegraphics[width=0.43\textwidth]{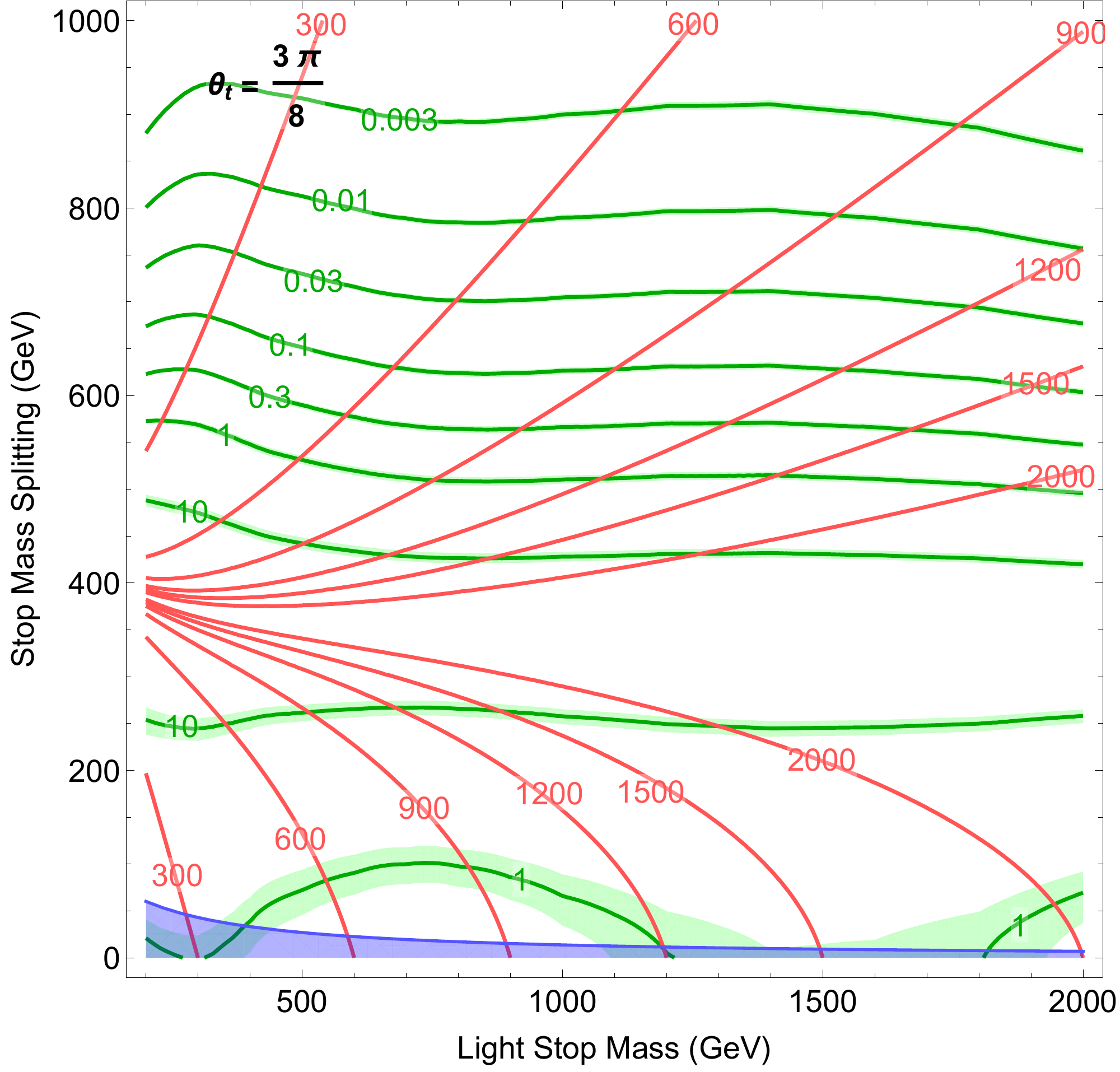}
\end{tabular}
\end{center}
\caption{\label{f.fsusy}
Green contours show $R_{\text{FSUSY}}$, conservatively estimating the number of glueballs produced in stop pair production normalized to exotic Higgs decays in FSUSY 
as a function of lightest stop 
mass and mass splitting, for purely RH light stop (top) and some mixing (bottom). Green shading shows the effect of varying the glueball mass $m_0$ from 15 GeV (right edge of band) to 50 GeV (left edge). Red contours show $m_\mathrm{eff}$, which corresponds to the left vertical axis in \fref{qlh}. Blue shading indicates where $\tilde b_L$ is lighter than $\tilde t_1$, allowing 
$\beta$ decay.
}
\end{figure}

\textbf{\emph{Signal Estimate ---}}
It is now straightforward 
 to estimate the number of $0^{++}$ glueballs produced in each top partner pair production event. In \cite{futurequirk} we will use this formalism to explore the landscape of possible quirk signals in detail. Here we 
  motivate that study by comparing the number of produced glueballs in top partner pair production to exotic Higgs decays, as discussed in \cite{Curtin:2015fna}. 

We assume 
the number of glueballs produced in the annihilation of two 62 GeV stops is the same as the number of glueballs produced in the decay of the 125 GeV Higgs boson. In computing the ratio 
of $0^{++}$ glueballs in the two processes, $r_{G_0}$ is about the same and drops out. We 
compare the signal rates by computing the ratio
\begin{equation}
\label{e.Rfsusy}
R_{\text{FSUSY}}=\frac{\sigma_\text{DY+VBF}(pp\to\tilde{t}_1\tilde{t}_1) \mathrm{Br}(\eta_{\tilde{t}}\to g_\text{B}g_\text{B}) N_G(2 m_{\tilde t_1})
} {\sigma_\text{VBF}(pp\to h)\varepsilon_\text{VBF}\text{Br}(h\to g_\text{B}g_\text{B})}
\end{equation}
where $N_G(125\gev)$ is normalized to 1, giving $\sim 4$ for $m_{\tilde t_1} = 2 \tev$. We have assumed VBF Higgs production, and $\varepsilon_\text{VBF} \approx 20\%$ is a generous estimate of the acceptance for VBF triggers \cite{Curtin:2015fna}. 
Detector efficiencies were considered in \cite{Curtin:2015fna} and roughly drop out of the ratio if detection of displaced decays is the primary discovery channel. (Computing the sensitivity of prompt searches to the production of multiple glueballs with sub-cm decay lengths requires the more careful treatment of the glueball momentum distribution 
in~\cite{futurequirk}.)

%
%

 This ratio is shown as the green contours in \fref{fsusy} for two stop mixing angles. The 
  large regions where this ratio is larger than 1 
  indicate more displaced vertices from top partner pair production than exotic Higgs decays. In fact, given 
  our conservative estimate of $N_G^0$, pair production is likely to be the superior discovery channel even when $R_{\text{FSUSY}}$ is somewhat smaller than 1. To understand the gain in top partner mass reach, we also show contours of $m_\mathrm{eff}$ (red), which corresponds to the left vertical axis of \fref{qlh}. The bounds on $m_\mathrm{eff}$ from exotic Higgs decays are $\sim 1 \tev$ at the HL-LHC, and a factor of 10 in signal corresponds to $\sim 200 \gev$ in reach. For unmixed RH stops (top plot), pair production is the discovery channel for masses $< 500 - 1000 \gev$.  Pair production is even more important for mixed stops (bottom plot), where exotic Higgs decays are suppressed by cancellations. In fact, for the moderately mixed example shown, quirky pair production is competitive or dominant for all $m_{\tilde t_1} < 2 \tev$. 
In either case, 
 the large glueball rate suggests 
  that top partner pair production will help probe the sub-cm glueball regime. 
 Note, however, that the annihilation branching fraction to mirror gluons becomes small for large mass splittings. In that case, 
  di-higgs searches may have greater sensitivity. 
For purely LH stops, the quirk state is likely to $\beta$-decay to mirror-sbottoms, 
the resulting leptons increasing the conspicuousness of the signal.


\textbf{QUIRKY LITTLE HIGGS --- }
The QLH model 
features a vector-like fermion top partner, which is an $SU(2)_L$ singlet with mass $m_T$ and 
 hypercharge $2/3$. 
A lower bound on the signal is 
 estimated as in 
  FSUSY, 
  with a few modifications. There is no mass splitting between different top partner states, allowing us to plot results in the same $(m_0, m_T)$ plane as the exotic Higgs decay bounds. 
VBF production is not competitive with DY and is 
omitted. 

One complication is that the quirks can 
annihilate as either a spin singlet ${}^1S_0$ (which can annihilate to di-gluons) or triplet ${}^3S_1$ (which annihilates to at least three gluons). The relevant 
annihilation widths are 
adapted from \cite{Barger:1987xg} by 
noticing that the quirks 
 do not receive most of their mass from the light Higgs VEV and do not couple axially to the $Z$-boson. 
We apply the same assumptions used to derive \eref{Rfsusy} to the QLH case, assuming annihilation dominantly through the ${}^1S_0$ state:
\begin{equation}
\label{e.Rqlh}
R_{\text{QLH}}=\frac{\sigma_\text{DY}(pp\to T\overline{T}) \mathrm{Br}({}^1S_0, {}^3S_1 \to g_\text{B}g_\text{B}(g_\text{B})) N_G(2 m_T)} {\sigma_\text{VBF}(pp\to h)\varepsilon_\text{VBF}\mathrm{Br}(h\to g_\text{B}g_\text{B})}
\end{equation}
The 
 peculiarities of fermionic quirk annihilation
may change the true value of this ratio by a factor of $\sim 2$, 
 but since $R_{\text{QLH}}$ likely represents an extreme under-estimate of the displaced signal detection rate, we ignore them for simplicity. 
(We have checked that dileptons from triplet annihilation are a less sensitive probe than displaced glueball decays \cite{atlasdilepton}.)

$R_{\text{QLH}}$ is shown as green contours overlaid on the 
projected Exotic Higgs decay bounds in \fref{qlh}. Note the top quirk mass 
is on the right vertical axis of that plot. We expect quirk annihilation to yield more signal events than exotic Higgs decays in the entire region of parameter space where the latter have sensitivity. Furthermore, as explained above, quirk annihilation may be the only reliable way of probing 
sub-cm glueball decay lengths.
This makes quirk 
pair production the main discovery channel for NN in the QLH scenario at the LHC.

\textbf{CONCLUSIONS ---}
This letter 
analyzes top partner pair production in theories of Neutral Naturalness at the LHC. This is particularly motivated for 
 top partners with EW charge  like Folded SUSY or the Quirky Little Higgs. In minimal models, mirror glueballs are 
  the bottom of the mirror spectrum, and the top partners form 
  quirky bound states which annihilate into jets of mirror gluons. The unknown details of mirror hadronization 
  are parameterized 
  in a way that is transparent, allows for $\mathcal{O}(1)$ signal estimates,  
  can be applied consistently event-by-event, and factorizes from perturbative theory parameters like the top partner mass. 

Our analysis shows that production of mirror glueballs in top partner pair production, which can give rise to displaced decay signals or high multiplicities of hard $\bar b b$ and $\tau^+ \tau^-$ pairs at the LHC, can be competitive or dominant to glueball production in exotic Higgs decays as analyzed in \cite{Curtin:2015fna, Csaki:2015fba}. Furthermore, it may be the only reliable way to experimentally access 
glueball lifetimes 
below a cm where prompt searches might suffer significant backgrounds. Consequently, 
top partner pair production is the likely discovery channel of NN in the QLH model, and 
many FSUSY scenarios.

The landscape of signatures obtained from 
top partner pair production is rich, sharing some qualitative
 features with the 
 Emerging Jets scenario \cite{Schwaller:2015gea}. 
 A particularly tantalizing possibility is to measure the top partner masses and couplings directly to ascertain if 
 the NN mechanism solves 
 the Little Hierarchy Problem.

\acknowledgments

\textbf{Acknowledgements:} 
We thank Markus Luty, Yuhsin Tsai, George Sterman, and the participants of the 2015 CERN-CKC Neutral Naturalness workshop for useful discussion. We thank Olivier Mattelaer for helping us with MadGraph. Z.C., D.C., and C.B.V. are supported by National Science Foundation grant No. PHY-1315155 and the Maryland Center for Fundamental Physics.

\bibliography{detectcolorlesstops}

\end{document}